\documentclass[superscriptaddress,amsmath,amssymb,twocolumn]{revtex4-1}

\pdfoutput=1

\usepackage{graphicx}
\usepackage{chemarr}
\usepackage{times,longtable}
\usepackage{hyperref}
\usepackage{color}
\usepackage{afterpage}
\usepackage{soul,comment}
\usepackage[usenames,dvipsnames]{xcolor}
\usepackage{colortbl}
\usepackage{pbox}
\usepackage{capt-of}% or use the larger `caption` package

\graphicspath{{large/}}

\bibpunct{[}{]}{,}{n}{}{,}

\newcommand{\inn}{\textrm{in}}
\newcommand{\out}{\textrm{out}}
\newcommand{\opt}{\textrm{opt}}
\newcommand{\TF}{\textrm{TF}}
\newcommand{\mirna}{\textrm{miRNA}}

\newcommand{\avg}[1]{\left\langle #1 \right\rangle}
\renewcommand{\bar}[1]{\overline{#1}}

\usepackage{chemarr}
\usepackage{epsf,mathtools}
\usepackage{graphicx}% Include figure files
\usepackage{dcolumn}% Align table columns on decimal point
\usepackage{bm}% bold math
\begin{document}

%\title{Competition and effective RNA-RNA interactions in minimal models of Post-Transcriptional Regulatory Networks}

\title{Probing the Limits to MicroRNA-Mediated Control of Gene Expression}

\author{Araks Martirosyan}
\affiliation{Dipartimento di Fisica, Sapienza Universit\`a di Roma, Rome, Italy}
\affiliation{Soft and Living Matter Lab, Institute of Nanotechnology (CNR-NANOTEC), Consiglio Nazionale delle Ricerche, Rome, Italy}

\author{Matteo Figliuzzi}
\affiliation{Sorbonne Universit\'es, UPMC, Institut de Calcul et de la Simulation, Paris, France}
\affiliation{Sorbonne Universit\'es, UPMC, UMR 7238, Computational and Quantitative Biology, Paris, France}
\affiliation{CNRS, UMR 7238, Computational and Quantitative Biology, Paris, France}

\author{Enzo Marinari}
\thanks{Authors contributed equally}
\affiliation{Dipartimento di Fisica, Sapienza Universit\`a di Roma, Rome, Italy}
\affiliation{INFN, Sezione di Roma 1, Rome, Italy}

\author{Andrea De Martino}
\thanks{Authors contributed equally}
\affiliation{Dipartimento di Fisica, Sapienza Universit\`a di Roma, Rome, Italy}
\affiliation{Soft and Living Matter Lab, Institute of Nanotechnology (CNR-NANOTEC), Consiglio Nazionale delle Ricerche, Rome, Italy}
\affiliation{Center for Life Nano Science@Sapienza, Istituto Italiano di Tecnologia, Rome, Italy}

\begin{abstract}
{\bf Abstract --} According to the `ceRNA hypothesis', microRNAs (miRNAs) may act as mediators of an effective positive interaction between long coding or non-coding RNA molecules, carrying significant potential implications for a variety of biological processes. Here, inspired by recent work providing a quantitative description of small regulatory elements as information-conveying channels, we characterize the effectiveness of miRNA-mediated regulation in terms of the optimal information flow achievable between modulator (transcription factors) and target nodes (long RNAs). Our findings show that, while a sufficiently large degree of target derepression is needed to activate miRNA-mediated transmission, (a) in case of differential mechanisms of complex processing and/or transcriptional capabilities, regulation by a post-transcriptional miRNA-channel can outperform that achieved through direct transcriptional control; moreover, (b) in the presence of large populations of weakly interacting miRNA molecules the extra noise coming from titration disappears, allowing the miRNA-channel to process information as effectively as the direct channel. These observations establish the limits of miRNA-mediated post-transcriptional cross-talk and suggest that, besides providing a degree of noise buffering, this type of control may be effectively employed in cells both as a failsafe mechanism and as a preferential fine tuner of gene expression, pointing to the specific situations in which each of these functionalities is maximized. \\
~\\
{\bf Author Summary --} The discovery of RNA interference has revolutionized the decades' old view of RNAs as mere intermediaries between DNA and proteins in the gene expression workflow. MicroRNAs (or miRNAs), in particular, have been shown to be able to both stabilize the protein output by buffering transcriptional noise and to create an effective positive interaction between the levels of their target RNAs through a simple competition mechanism known as `ceRNA effect'. With miRNAs commonly targeting multiple species of RNAs, the potential implication is that RNAs could regulate each other through extended miRNA-mediated interaction networks. Such cross-talk is certainly active in many specific cases (like cell differentiation), but it's unclear whether the degree of regulation of gene expression achievable through post-transcriptional miRNA-mediated coupling can effectively overcome the one obtained through other mechanisms, e.g. by direct transcriptional control via DNA-binding factors. This work quantifies the maximal post-transcriptional regulatory power achievable by miRNA-mediated cross-talk, characterizing the circumstances in which indirect control outperforms direct one. The emerging scenario suggests that, in addition to its widely recognized noise-buffering role, miRNA-mediated control may indeed act as a master regulator of gene expression.
\end{abstract}

\maketitle

\section*{Introduction}

The problem of tuning protein expression levels is central for eukaryotic cell functionality. A variety of molecular mechanisms are implemented to guarantee, on one hand, that protein copy numbers stay within a range that is optimal in the given conditions and, on the other, that shifts in expression levels can be achieved efficiently whenever necessary  \cite{LopezMaury,Guil,Cech}  (whereby `efficiency' here encompasses both a dynamical characterization, in terms of the times required to shift, and a static one, in terms of moving as precisely as possible from one functional range to another). Quantifying and comparing their effectiveness in different conditions is an important step to both deepen our fundamental understanding of regulatory circuits and to get case-by-case functional insight about why a specific biochemical  network has been selected over the others.

As the major direct regulators of gene expression, transcription factors (TFs) are most immediately identified as the key potential modulators of protein levels \cite{Latchman}. In a somewhat simplified picture, one may imagine that a change in amount of a TF can induce a change in the expression level of the corresponding gene, and that the ability to regulate the latter (the output node) via the former (the input node) can be assessed by how strongly the two levels correlate. The effectiveness of a regulatory element is however limited by the stochasticity of intracellular processes, from the TF-DNA binding dynamics to translation \cite{Raser}. A convenient framework to analyze how noise constrains regulation is provided by information theory \cite{Shannon1,Shannon2}. In particular, the simplest situation in which a single TF modulates the expression of a single protein can be characterized analytically under the assumption that the noise affecting the input-output channel is sufficiently small. 
 The mutual information between modulator and target --a convenient quantity through which regulatory effectiveness can be characterized-- depends on the distribution of modulator levels and can be maximized over it. Remarkably, in at least one case this maximum has been found to be almost saturated by the actual information flow measured in a living system (for more details see \cite{Callan,Bialek}). In other terms, for sufficiently small noise levels in the channel that links TFs to their functional products, one may quantify the optimal regulatory performance achievable in terms of the maximum number of bits of mutual information that can be exchanged between modulator and target.
 
Several control mechanisms however act at the post-transcriptional level \cite{Fire,Mello,Baker}. Among these, regulation by small regulatory RNAs like eukaryotic microRNAs (miRNAs) has attracted considerable attention over the past few years \cite{Bartel,Valencia,Chekulaeva}. In short, miRNAs are small non-coding RNA molecules encoded by nuclear DNA, that can inhibit  translation or catalyze degradation of mRNAs when bound to them via protein-mediated base-pairing.  miRNAs appear to be crucial in an increasing number of situations ranging from development to disease \cite{Hwang,Lynn,Shi}. Their function however can differ significantly from case to case. For instance, they have been well characterized as noise buffering agents in protein expression \cite{Osella,Siciliano} or as key signaling molecules in stress response \cite{Leung}. Recently, though, investigations of the miRNA-mediated post-transcriptional regulatory (PTR) network have hinted at a possibly more subtle and complex role. It is indeed now clear \cite{Lee,Lewis,Selbach,Baek,BartelDP,Rhoades} that the miRNA-RNA network describing the potential couplings stretches across a major fraction of the transcriptome, with a large heterogeneity both in the number of miRNA targets and in the number of miRNA regulators for a given mRNA. The competition effects that may emerge in such conditions suggest that miRNAs may act as channels through which perturbations in the levels of one RNA could be transmitted to other RNA species sharing the same miRNA regulator(s). Such a scenario has been termed the `ceRNA effect', whereby ceRNA stands for `competing endogenous RNA' \cite{Salmena}. In view of its considerable regulatory and therapeutic implications, the ceRNA effect has been extensively analyzed both theoretically and experimentally \cite {Jens,Denzler,Cesana,Karreth,Ebert,Bosson,Memczak,Levine, Mehta,Figliuzzi,Bosia,CarlaB,MFigliuzzi,Yuan,Chiu,Tay}.

The apparent ubiquity of potentially cross-talking ceRNAs however raises a number of fundamental questions about the effectiveness of ``regulation via competition'' {\it per se}. Although hundreds of targets are predicted for a single miRNA, observations show that only few of them are sensitive to  changes in miRNA expression levels. Most targets are likely to provide a global buffering mechanism through which miRNA levels are overall stabilized \cite{Salmena,Jens}. Effective competition between miRNA targets requires that the ratio of miRNA molecules to the number of target sites lies in a specific range, so that the relative abundance of miRNA and RNA species must be tightly regulated for the ceRNA mechanism to operate \cite{Yuan,Chiu,Tay,Bosia,Jens,Denzler}. On the other hand, the magnitude of the ceRNA effect is tunable by the miRNA binding and mRNA loss rates \cite{Yuan,Bosia,Figliuzzi}. The performance of a regulatory element, however, does not only depend on kinetic parameters, but also on the range of variability (and possibly on the distribution itself) of modulator levels (e.g. TFs) \cite{Callan,Bialek}. The maximal regulatory effectiveness of a given genetic circuit --quantifying how precisely the output level can be determined by the input level-- can therefore generically be obtained by solving an optimization problem over the distribution of inputs. This type of approach provides an upper bound to the effectiveness of a regulatory mechanism as well as indications concerning which parameters, noise sources and/or interactions most hamper its performance.

It would be especially important to understand in which conditions the degree of control of the output variable (i.e. the ceRNA/protein level) that can be accomplished through post-transcriptional miRNA-mediated cross-talk may exceed that obtainable by different regulatory mechanisms. In this work we characterize the maximal regulatory power achievable by miRNA-mediated control and compare it with that of a direct, TF-based transcriptional unit \cite{Tkacik}. In principle, since fluctuations can be reduced by increasing the number of molecules, an (almost) arbitrary amount of information can be transmitted through a biochemical network. However, cells have to face the burden of macromolecular synthesis \cite{TkacikCallan,Dekel,Tanase}. Optimality is therefore the result of a trade-off between the benefits of reduced fluctuations and the drawbacks of the associated metabolic costs. For this reason, we start by fixing a maximal rate of transcription (or, alternatively, the maximal number of output molecules) so as to have a simple but reasonable framework to characterize and compare the capacities of the different regulatory channels. Next, we quantify how an input signal is processed by the transcriptional (TF-based) and post-transcriptional (miRNA-mediated) regulatory elements by characterizing the response in the output ceRNA's expression levels. In such a setting, information flow is hampered by intrinsic noise if the target gene is weakly derepressed by the activation of its competitor. Otherwise, target derepression appears to have a strong impact on a regulatory element's capacity. Upon varying the magnitude of derepression by tuning the kinetic parameters, we then show  that in certain regimes miRNA-mediated regulation can indeed outperform direct control of gene expression. Finally, we argue that the presence of miRNA molecules in large copy numbers notably reduces the level of intrinsic noise on weakly targeted transcripts. In this case, the mutual regulation of ceRNA molecules by miRNA-mediated channels may become a primary mechanism to finely tune gene expression.

Besides providing a quantitative characterization of the maximal regulatory power achievable through miRNA-based post-transcriptional control, these results provide important hints on the circumstances in which regulation by small RNAs may function as the main tuner of gene expression in cells.

\section*{Results}

\subsection*{Mathematical model of ceRNA competition}

\begin{figure}
\begin{center}
\includegraphics[width=8.5cm]{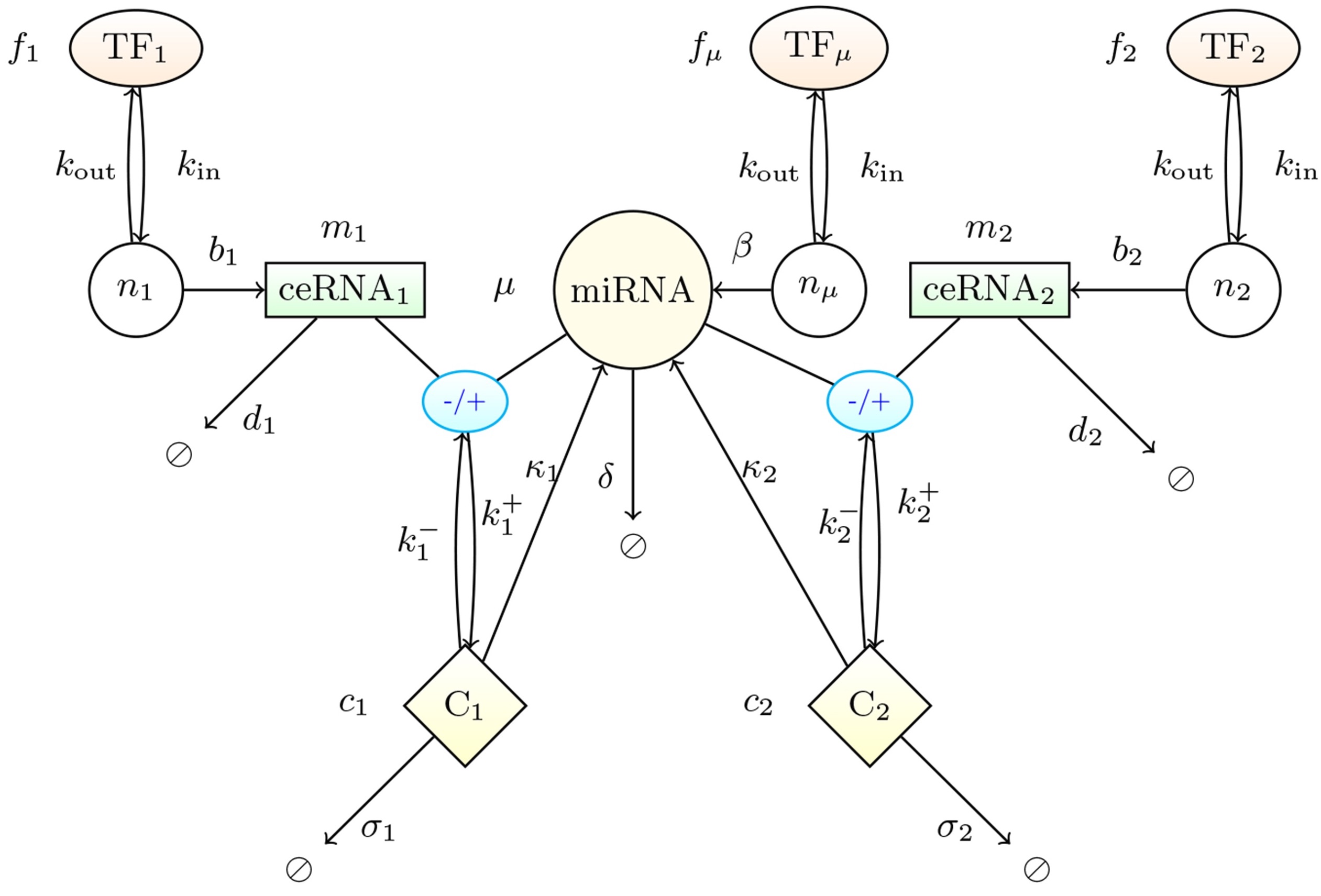}
\end{center}
\caption{ {\bf Schematic representation of the basic ceRNA network model.} The model comprises three transcription factors (TF$_1$, TF$_2$ and TF$_\mu$) controlling the synthesis of two ceRNA species (ceRNA$_1$ and ceRNA$_2$) and one miRNA species, respectively. The fractional occupancies of TF binding sites are denoted by $n_1$, $n_2$ and $n_\mu$, respectively. Both ceRNAs are targets for the miRNA, with whom they form complexes denoted respectively as C$_1$ and C$_2$. The amount of molecules for each species is denoted by the variable next to the corresponding node. Reaction rates are reported next to the corresponding arrow.
\label{Fig1}}
\end{figure}

We consider (see Fig. \ref{Fig1}) a system formed by two ceRNA species (whose levels are labeled $m_1$ and $m_2$, respectively) and one miRNA species (with level labeled $\mu$), whose transcriptions are activated by a single TF each (with levels labeled, respectively, $f_1$, $f_2$ and $f_\mu$). Both ceRNAs are in turn targeted by the miRNA. miRNA-ceRNA complexes (levels labeled $c_i$ with $i=1,2$) assemble and disassemble at rates $k_i^\pm$, respectively, whereas complexes can be degraded both stoichiometrically (i.e. without miRNA recycling) at rates $\sigma_i$ and catalytically (i.e. with miRNA recycling) at rates $\kappa_i$ \cite{Muers}. In addition, ceRNA and miRNA molecules degrade (resp. synthesize) at rates $d_i$ (resp. $b_i$) and $\delta$ (resp. $\beta$), respectively. Steps leading to the formation of the RNA-induced silencing complex (RISC), allowing for the miRNA-ceRNA binding, are neglected for simplicity. TF levels are treated as externally controlled parameters.

Denoting the TF-DNA binding/unbinding rates by $k_{\inn}$ and $k_{\out}$, respectively (for simplicity, these parameters are taken to be the same for all involved TFs),  the TF binding sites' fractional occupancies $n_{\ell}$ ($0\leq n_{\ell}\leq 1$, $\ell\in\{1,2,\mu\}$) obey the dynamics
\begin{equation} 
\frac{d n_{\ell}}{dt} = k_{\inn}(1-n_{\ell}) f_\ell^h - k_{\out} n_{\ell}~~,
\end{equation}
according to which transcriptional activation requires the cooperative binding of $h$ TF molecules for each RNA species involved.  In general, the occupancies by different TFs equilibrate on different timescales \cite{GT}. However  it is often assumed that the transcriptional on/off dynamics is much faster than transcription itself \cite{Alon,review}. As a consequence, each $n_{\ell}$ can be fixed at its `equilibrium' value.

\begin{equation}
\label{eq:nll}
\bar{n}_{\ell} = \frac{ k_{\inn} f_\ell^h } {k_{\inn} f_\ell^h + k_{\out} }~~.
\end{equation}

Every process in the above scheme contributes to the overall level of noise. We represent the mass-action kinetics of the system through the set of coupled Langevin processes ($i=1,2$)  \cite{Kampen,Mehta}
\begin{widetext}
\begin{gather}
\frac{d m_i }{dt}= -d_i m_i + b_i \bar{n}_{i} - k^{+}_{i} \mu m_i + k^- c_i + 
\xi_{i} - \xi^+_i + \xi_i^- ~~, \nonumber\\
\frac{d c_i }{dt} = k^+_i \mu m_i - (k_i^- + \kappa_i+\sigma_i) c_i + \xi_{i}^{\sigma} + \xi^+ _i - \xi^-_i - \xi^\kappa_i ~~, \label{eq:SDEeqs} \\
\frac{d \mu}{dt} = - \delta \mu + \beta \bar{n}_{\mu}  - \sum_i k_i^{+} \mu m_i + \sum_i ( k^-_i + \kappa_i) c_i
 + \xi_{\mu}- \sum_i \xi^+_i + \sum_i \xi^-_i + \sum_i \xi^\kappa_i~~,\nonumber
\end{gather}
\end{widetext}
where the mutually independent random (Poisson) `forces' $\xi_i$, $\xi_i^\pm$, $\xi_\mu$, $\xi_i^\kappa$ and $\xi_{i}^\sigma$ denote, respectively, the intrinsic noise in ceRNA levels (due to random synthesis and degradation events), in the association/dissociation processes of complexes, in the miRNA level, in the catalytic complex decay and in the stoichiometric complex decay. Each of the above noise terms has zero mean, while correlations are given by

\begin{gather}
\avg{\xi_{i}(t)\xi_{i}(t')} =  (d_i \bar{m}_i + b_i\bar{ n}_{i} ) ~ \delta(t-t')~~,\\
\avg{\xi_{i}^+(t)\xi_i^+(t')} =  k_i^+ \bar{ m}_i \bar{\mu} ~ \delta(t-t')~~,\\
\avg{\xi_{i}^-(t)\xi_i^-(t')} =  k_i^- \bar{ c}_i ~ \delta(t-t')~~,\\
\avg{\xi_{\mu}(t)\xi_{\mu}(t')} = (\delta \bar{\mu} + \beta \bar{n}_\mu ) ~ \delta(t-t')~~, \\
\avg{\xi_{i}^\kappa(t)\xi_i^\kappa(t')} =  \kappa_i \bar{ c}_i ~ \delta(t-t')~~,\\
\avg{\xi_{i}^\sigma(t)\xi_{i}^\sigma(t')} = \sigma_i \bar{c}_i ~ \delta(t-t')~~,
\end{gather}
where we introduced the steady state molecule numbers 

\begin{gather}
\label{eq:steadystateM}
\bar{m}_i = \frac{b_i \bar{n}_{i} + k_i^- \bar{c}_i }{d_i + k_i^+ \bar{\mu} }~~, \\ 
\bar{\mu} = \frac{\beta \bar{n}_\mu + \sum_i (k_i^- + \kappa_i) \bar{c}_i }{ \delta + \sum_i k^+_i \bar{m}_i }~~,\\
\bar{c}_i = \frac{k^+_i \bar{\mu} \,\, \bar{m}_i }{\sigma_i + k^-_i + \kappa_i}~~.
\label{eq:steadystateMEND}
\end{gather}

\begin{figure}
\begin{center}
\includegraphics[width=8.5cm]{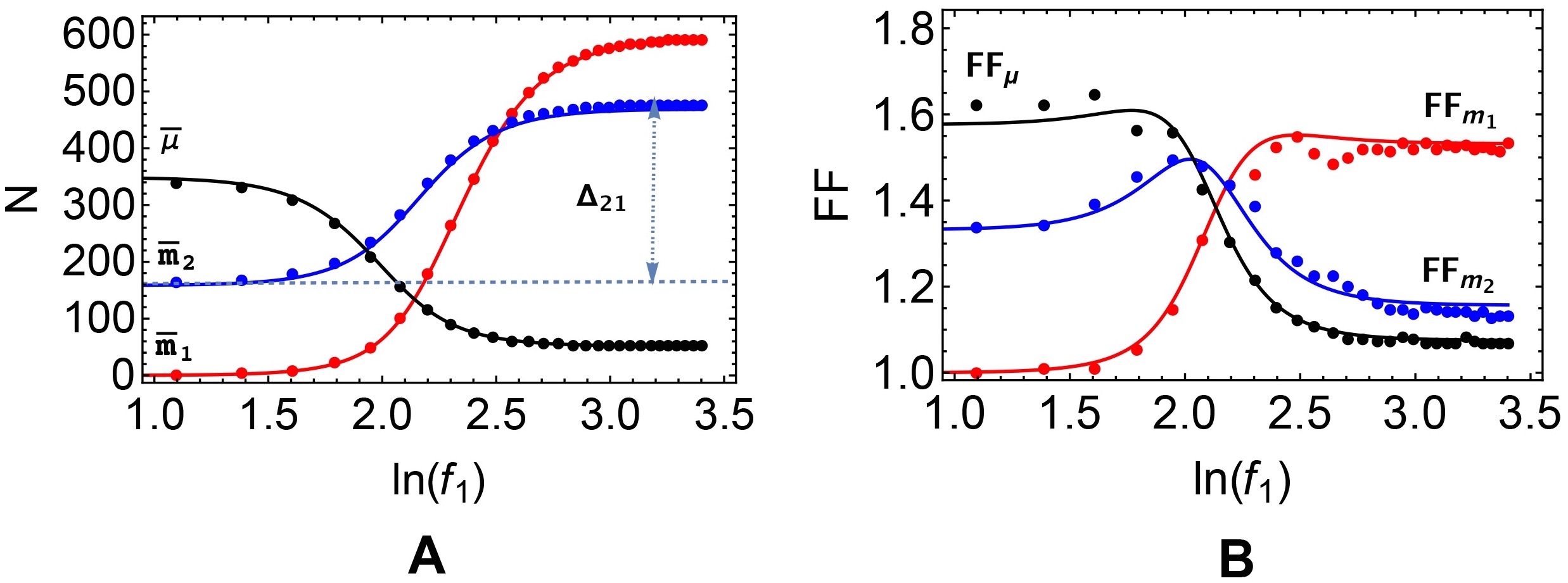}
\end{center}
\caption{
\label{Fig2}
{\bf Cross-talk scenario in the ceRNA network.}
Steady state values of {\bf (A)} the ceRNA and miRNA levels, and {\bf (B)} the corresponding Fano factors. Markers denote GA computations, curves correspond to approximate analytical solutions obtained by the linear noise approximation. Values of the kinetic parameters are reported in Table \ref{table1}. 
}
\end{figure}

We shall be interested in the fluctuations of molecular levels around the steady state that arise due to intrinsic noise sources. The stochastic dynamics of the system can be simulated via the \nameref{GA} (GA, see \nameref{MM}). Fig. \ref{Fig2}A shows typical GA results for $\bar{m}_1$, $\bar{m}_2$ and $\bar{\mu}$, with the corresponding Fano factors (FFs) shown in Fig. \ref{Fig2}B, as functions of $f_1$. Numerical results are matched against analytical estimates obtained by the linear noise approximation (see \nameref{MM}). 

In particular, in Fig. \ref{Fig2} one sees that an increase of $\bar{m}_1$ is accompanied by a concomitant increase of $\bar{m}_2$ and by a decrease of the average number of free miRNA molecules. This is an instance of miRNA-mediated ceRNA cross-talk. Indeed, upon up-regulating $m_1$ by injecting the corresponding TF (i.e. by increasing $f_1$), the level of free miRNAs will decrease as more and more molecules will be actively repressing ceRNA$_1$, causing in turn an up-regulation of ceRNA$_2$. $f_1$ will thus positively correlate with $m_2$. One easily sees that steady-state ceRNA levels depend on $\bar{\mu}$ through a sigmoidal function, namely  \cite{Figliuzzi}
\begin{equation}
\bar{m}_i = \frac{b_i \bar{n}_{i}} {d_i} F_i[\bar{\mu}]~~~~~,
~~~~~F_i[\bar{\mu}] = \frac{\mu_{0,i}}{\mu_{0,i} + \bar{\mu}}~~.
\end{equation}
The constant $\mu_{0,i} = \frac{d_i}{k_i^+} (1 +\frac{k_i^-}{\sigma_i + \kappa_i})$ acts as a `soft' threshold for the miRNA level, allowing to distinguish three situations:
\begin{itemize} 
\item if $\bar\mu \ll \mu_{0,i}$, ceRNA$_i$ is \emph{free} or {\it unrepressed}: spontaneous ceRNA degradation dominates over miRNA-mediated decay channels, so that, effectively, the ceRNA level is weakly sensitive to small changes in $\bar{\mu}$;
\item if $\bar\mu \gg \mu_{0,i}$, ceRNA$_i$ is \emph{bound} or {\it repressed}: miRNA-mediated ceRNA decay dominates over spontaneous ceRNA degradation but, again, the ceRNA level is weakly sensitive to small changes in $\bar{\mu}$ as most ceRNAs are bound in complexes with the miRNA;
\item if  $\bar\mu \simeq \mu_{0,i}$, ceRNA$_i$ is \emph{susceptible} to $\mu$: spontaneous and miRNA-mediated decay channels have comparable weights and the ceRNA level is very sentitive to small changes in $\bar{\mu}$. 
\end{itemize} 
The behaviour of the FFs (see Fig. \ref{Fig2}B) emphasizes  how noise patterns change in the different regimes. 
In the displayed example, ceRNA$_1$ and ceRNA$_2$ become susceptible for $\ln (f_1) \simeq 2.5 $ and $\ln (f_1) \simeq 2.1$, respectively. Indeed, one observes that the corresponding FFs peak close to these values, in accordance with the observation that stochastic fluctuations are enhanced when the rates of substrate supply are adequately balanced in a stoichiometrically coupled system \cite{Bosia,Elf,Noorbakhsh}.  The FF for ceRNA$_2$ appears to approach one for large values of $f_1$, as expected for the pure Poisson birth/death process that characterizes the free regime \cite{Rash}.
 On the other hand, for very small but nonzero mean fractional occupancy $\bar{n}_1$ (corresponding to small values of $f_1$), $m_1$ will with high probability only take on the values 0 or 1, as a transcribed molecule will quickly undergo degradation or sequestration in a complex. In such a situation, the mean and variance of $m_1$ will be calculated by summing up zeros and ones over time, leading to a FF equal to one.

\subsection*{Measuring the performance of regulatory elements}

\subsubsection*{The amplitude of ceRNA derepression}

\begin{figure*}
\begin{center}
\includegraphics[width=17cm]{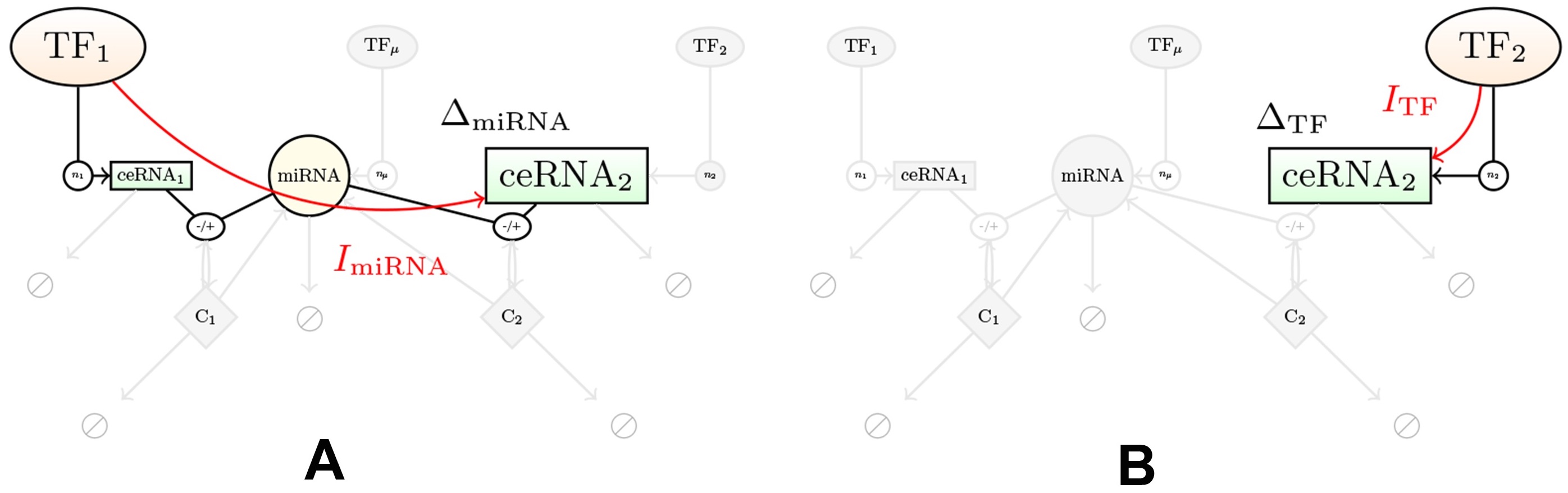}
\end{center}
\caption{
{ \bf Schematic representation of the channels under consideration.} 
\label{Fig3}
 {\bf (A)}  miRNA-channel, {\bf (B)} TF-channel. }
\end{figure*}

We shall focus on quantifying the regulatory capacity of the post-transcriptional and transcriptional channels that link, respectively, TF$_1$ to ceRNA$_2$ and TF$_2$ to ceRNA$_2$, as depicted in Fig. \ref{Fig3}.

In what follows, we shall take the number of transcription factors as the key input variables ($f_1$ for the miRNA-mediated channel and $f_2$ for the direct channel), and the number of target RNAs $m_2$ as the key output variable. It is reasonable to think that an effective regulator should have a sensible impact on the target molecule. As a first measure of regulatory performance we will therefore consider the \emph{amplitude of variation} (AOV) of $m_i$ via $f_j$, defined as
\begin{equation}
\Delta_{ij}=\bar{m}_i(f_j^{\max})-\bar{m}_i(f_j^{\min})~~,
\end{equation} 
corresponding to the difference between the maximum  and the minimum steady state level of ceRNA$_i$ obtained by varying $f_j$ from $f_j^{\min}$ to $f_j^{\max}$. Clearly, larger values of $\Delta_{ij}$ imply stronger degrees of derepression of ceRNA$_i$ induced by the corresponding TF. For sakes of simplicity, we shall adopt the following notation to distinguish the AOV induced by transciptional and miRNA-mediated control channels:

\begin{gather}
\Delta_{22}\equiv \Delta_\TF~~,\\
\Delta_{21}\equiv \Delta_\mirna~~.
\end{gather}
Note that the AOV only involves mean levels. As such, it is unable to account for the intrinsic degree of stochasticity of the system.

\subsubsection*{Maximal mutual information}

In order to evaluate the impact of noise, as a further quantifier of regulatory power we shall employ the mutual information (MI), a standard information-theoretic measure of channel performance given by ($i,j=1,2$) \cite{review} 
\begin{equation}
I(m_i, f_{j})=\int  p(m_i, f_{j}) \log_2 \frac{p(m_i, f_{j}) }{p(m_i) p(f_{j})}dm_i df_{j}~~,
\end{equation}
where $p(m_i, f_{j})$ stands for the joint probability distribution of $m_i$  and $f_j$, while $p(m_i)$ and $p(f_{j})$ denote its marginals. The MI is a non-negative quantity that vanishes when $m_i$ and $f_{j}$ are independent. Conversely, when $I(m_i, f_{j}) = \mathcal{I}$ one can reliably distinguish (roughly) $2^{\mathcal{I}}$ different values of the output variable upon tuning the input (see \cite{review,Tkacik,Bialek} for further details about the usefulness of this scheme for the analysis of transcriptional regulatory or biochemical modules).  The maximal achievable information flow for a fixed channel (i.e. for given kinetic parameters) is called the {\it channel capacity} and can be computed by maximising the MI over the input variable (in this case, over the distribution of the TF level). In the simplest case, one can assume that the channel is Gaussian, meaning that for every given input $f_j$ the output $m_2$ is described by the conditional probability density
\begin{equation}
p(m_2|f_{j}) = \frac{1}{\sqrt{2 \pi \sigma^2_{m_2}}} \exp\left[- \frac{(m_2 - \bar{m}_2(f_j))^2}{2 \sigma^2_{m_2}(f_{j})}\right]~~,
\end{equation}
where both the average $\bar{m}_2(f_j)$ and variance $\sigma_{m_2}^2 (f_j) = \avg{(m_2-\bar{m}_2(f_j))^2}$ depend on the input variable $f_j$.  In the limit of small variance, the distribution $p_{\opt}(f_{j})$ of TF levels that maximises the MI between $m_2$ and $f_j$ reads \cite{Tkacik}
\begin{equation}
\label{eq:popt}
p_{\opt}(f_{j}) = \frac{1}{Z} \left( \frac{1}{2 \pi e \sigma_{m_2}^2(f_j)} \left[\frac{ d \bar{m}_2 (f_j)}{ d f_{j}}\right]^2\right)^{1/2}~~,
\end{equation}
where
\begin{equation} \label{eq:zzeta}
Z = \int_{f_j^{\min}}^{f_j^{\max}} df_j \left(\frac{1}{2 \pi e \sigma_{m_2}^2(f_j)} \left[\frac{d\bar{m}_2 (f_{j}) }{ d f_{j}}\right]^2 \right)^{1/2}
\end{equation}
is a normalizing factor. For $p=p_\opt$ one obtains the channel capacity
\begin{equation}
I _{\opt}(m_2, f_j)= \log_2 Z~~.
\label{eq:Iopt}
\end{equation}
S1 Figure provides a qualitative yet intuitive representation of how the shape of the input/output relation and the size of fluctuations affect information flow.
Intuitively, the normalizing factor Z 'counts' the number of distinguishable output levels comparing the local slope of the input/output curve with the noise strength: if the slope is too low or the fluctuations are too high no information can be transmitted through the channel.
We shall use the following notation to distinguish the capacities of the direct and indirect channels:

\begin{gather}
I_\opt(m_2, f_2)\equiv I_\TF~~,\\
I_\opt(m_2, f_1)\equiv I_\mirna~~.
\end{gather}
We shall furthermore denote their difference by  $\Delta I = I_\TF- I_\mirna$.

\begin{figure}
\begin{center}
\includegraphics[width=8.5cm]{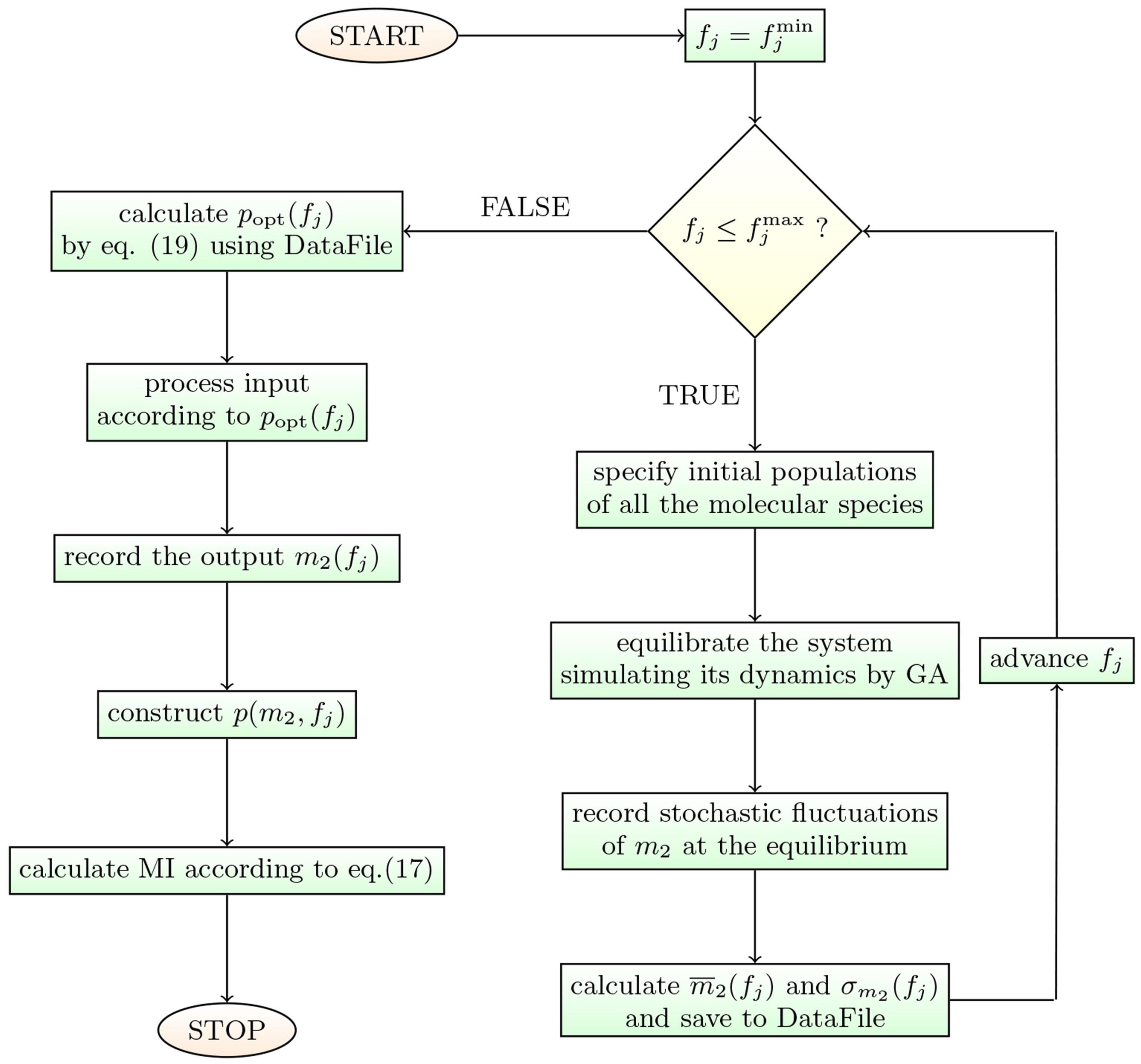}
\end{center}
\caption{
\label{Fig4}
{\bf Flowchart of the method.} A C++ implementation is available at \url{https://github.com/araksm/ceRNA}.
}
\end{figure}

We are henceforth going to adopt the following protocol. After generating the output data ($\bar{m}_2, \sigma_{m_2}$) by simulating the circuit's dynamics via the GA, the optimal input distribution will be constructed according to Eq. \eqref{eq:popt}. We shall then let the circuit process an input signal sampled from that distribution and calculate the MI for both the transcriptional and post-transcriptional channels (see the flowchart in Fig. \ref{Fig4}). Channel capacities  will then be compared, with the goal of clarifying the conditions under which the former can be more effective than the latter.

\subsection* {Comparing indirect miRNA-mediated regulation with direct transcriptional control}

Target prediction algorithms suggest that miRNA binding affinities vary significantly among their targets \cite{Breda}. Such heterogeneities indeed are mapped to the miRNA binding kinetics and are shown to influence the susceptibility of the targets to the miRNA molecules \cite{Bosson}. Moreover, the level of complementarity between the regulator and the target seems to be decisive for the selection of a decay channel (catalytic or stoichiometric) for the miRNA-ceRNA complex \cite{BartelDP,Valencia}. One may therefore expect that the effectiveness of miRNA-mediated post-transcriptional control depends strongly on the kinetic parameters characterizing the network.
 
Here in particular, we are going to investigate how the capacities of these regulatory elements are affected by changes in (i) miRNA-ceRNA binding kinetics, (ii) miRNA recycling rates, and (iii)  effective transcription rates of all RNA species involved.

In order to contrast the performances of miRNA- and TF-channels we shall start by analyzing their respective responses to the same input signal. 
More precisely, the input variable $f_j$ will be varied from 0 to a value $f^{\max}$ defined by the condition  $n_{j} (f^{\max}) = 0.99$ (i.e. from a situation in which the promoter is always free to one in which it is essentially always occupied).

\subsubsection* {The capacity of the miRNA-channel is maximal in a specific range of miRNA-ceRNA binding rates} 

\begin{figure*}
\begin{center}
\includegraphics[width=17cm]{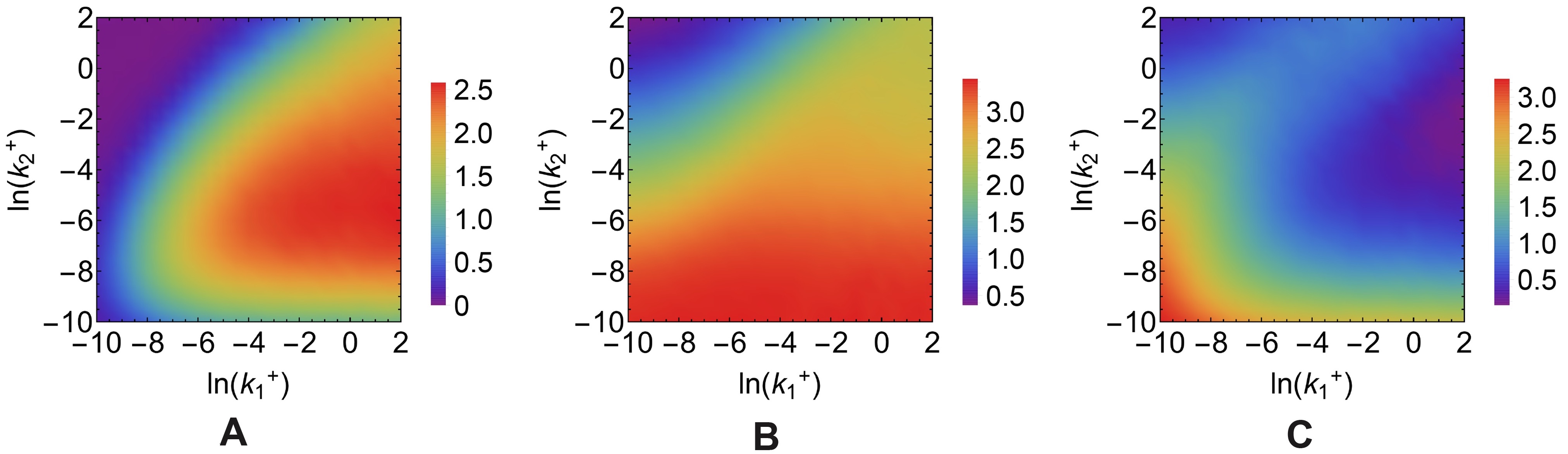}
\end{center}
\caption{
\label{Fig5}
{\bf Dependence of the transcriptional and post-transcriptional channel capacities on miRNA-ceRNA association rates.}
{\bf (A)} $I_\mirna$.
{\bf (B)} $I_\TF$
{\bf (C)} $\Delta I = I_\TF- I_\mirna$.
Values of the kinetic parameters are reported in Table \ref{table1}. 
}
\end{figure*}

Available estimates of RNA binding energetics indicate that the target sites of an individual  small non coding RNA can span a wide range of affinity both in mammalian cells \cite{Breda} and in bacteria \cite{Hao}. To explore the functional consequences of such heterogeneities we analyse the behaviour of the regulatory channels changing the values of the complex association rates  $k^+_1$ and $k^+_2$ while keeping fixed all others parameters. As displayed in Fig. \ref{Fig5}, the maximal amount of information transmitted through the post-transcriptional channel (Fig. \ref{Fig5}A) increases monotonically with $k_1^+$ (for fixed $k_2^+$) and displays a maximum versus $k_2^+$ (for fixed $k_1^+$). Indeed, the higher $k_1^+$, the higher $c_1$, and the more information about changes in $f_1$ can be transmitted through the miRNA-channel. On the other hand, ceRNA$_2$ is unrepressed by the miRNA for small values of $k_2^+$ while it is strongly repressed for high values of $k_2^+$, for any value of the input. In both regimes, ceRNA$_2$ is only weakly sensitive to changes in the level of its competitor. In such conditions, no information can be transmitted through the miRNA-channel. The information flow is therefore maximal between these two extremes. 

For the transcriptional channel, we expect to see maximum information transmission when the target, ceRNA$_2$, is unrepressed by the miRNA, in which case the channel is effectively decoupled from the rest of the network, while no information transmission is expected when the target is fully repressed by the miRNA. Indeed, in Fig. \ref{Fig5}B one sees that, for small values of $k_2^+$ (unrepressed target), $I_\TF$ is the highest and it decreases to zero as $k_2^+$ increases together with the degree of target repression. Moreover, since a higher $k_1^+$ moves $m_2$ to larger values (i.e. towards the free regime), $I_\TF$ increases with $k_1^+$ and decreases with $k_2^+$  in the susceptible regime. Notice that, when post-transcriptional capacity is highest, miRNA-mediated regulation is as effective as transcriptional control (see Fig. \ref{Fig5}C).

The intuition that the channel capacity is strongly coupled to the target's size of derepression (its AOV) is confirmed by the fact that the latter indeed displays a very similar behaviour (see S2 Figure). 

\subsubsection* {miRNA-mediated regulation may represent the sole control mechanism in case of differential complex processing}

\begin{figure*}
\begin{center}
\includegraphics[width=17cm]{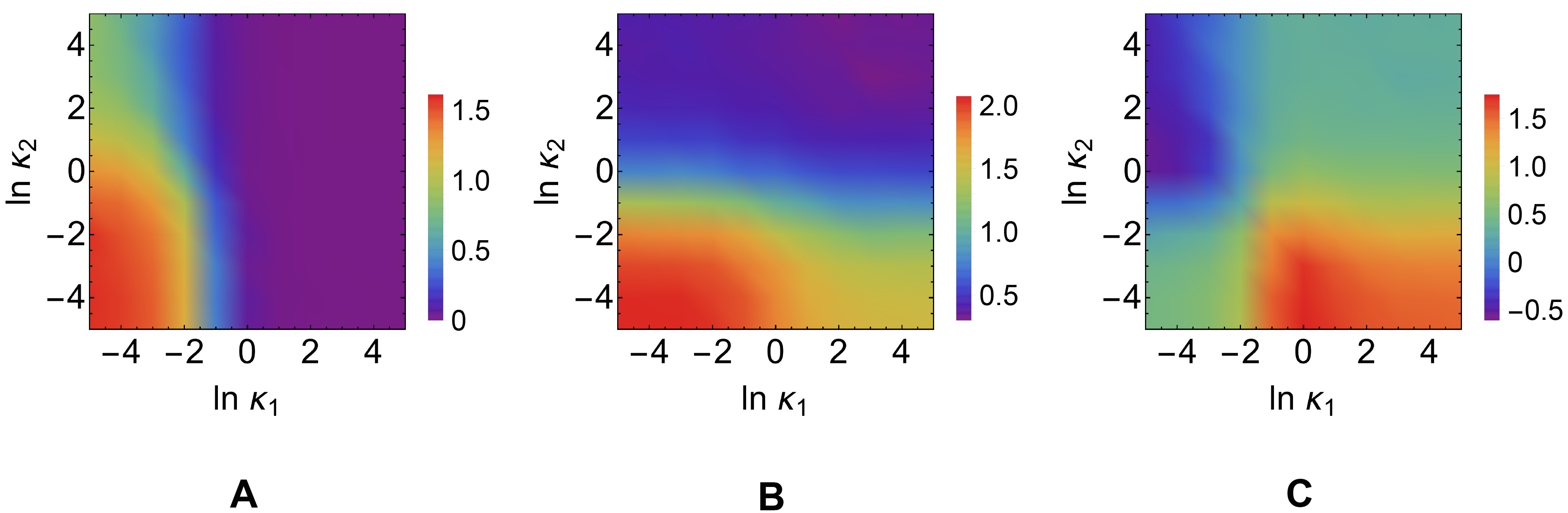}
\end{center}
\caption{
\label{Fig6}
{\bf Dependence of the transcriptional and post-transcriptional channel capacities on miRNA recycling rates.}
{\bf (A)} $I_\mirna$.
{\bf (B)} $I_\TF$
{\bf (C)}  $\Delta I = I_\TF- I_\mirna$.
Values of the kinetic parameters are reported in Table \ref{table1}. 
}
\end{figure*}

Direct experimental measurements of miRNA-recycling rates are challenging \cite{Baccarini}. However it is well known that the degreee of complementarity of a miRNA to its target lead to different repression pathways \cite{Hutvagner} and that differential complex processing strongly affects reciprocity of competition \cite{Yuan}. In order to better quantify the degree of regulatory control achievable in case of different processing mechanisms we changed the values of  miRNA-recycling rates $\kappa_1$ and $\kappa_2$ keeping fixed all the other parameters,  obtaining the results displayed in Fig. \ref{Fig6}.
Both $I_\TF$ and $I_\mirna$ are higher for small values of the recycling rates, and decrease as miRNA recycling increases. The observed dependence of $I_\mirna$ on $\kappa_1$ agrees with the results of \cite{Figliuzzi}, where the sensitivity of miRNA molecules to changes in the levels of their targets decreases as the miRNA recycling rate increases, eventually vanishing for $\kappa_1\gg \sigma_1$. Indeed, for large $\kappa_1$ miRNAs become insensitive to changes in $f_1$, and hence no information can be transmitted.

Remarkably (see Fig. \ref{Fig6}C), for large $\kappa_2$ and small $\kappa_1$ miRNA-mediated regulation may be the sole mechanism able to effectively control the target level. In this case, in fact, $I_{\TF}$ becomes especially small as the target is completely repressed by the miRNA while $I_{\mirna}$ stays finite since the activation of ceRNA$_1$ can titrate away the miRNA and therefore derepress $m_2$. 

As for the previous case, the behaviour of the AOV correlates strongly with this scenario, i.e. larger AOV corresponds to higher achievable information flow in the regulatory element (see S2 Figure). This suggests that, although targets that are degraded purely catalytically cannot compete for miRNA at steady state (as also shown for bacterial small RNAs \cite{Feng}), they can be notably derepressed if one of their competitors decays stoichiometrically.

\subsubsection*{Effective derepression of a repressed target may be conditional on the activation of its competitor and of miRNA}

\begin{figure}
\begin{center}
\includegraphics[width=8.5cm]{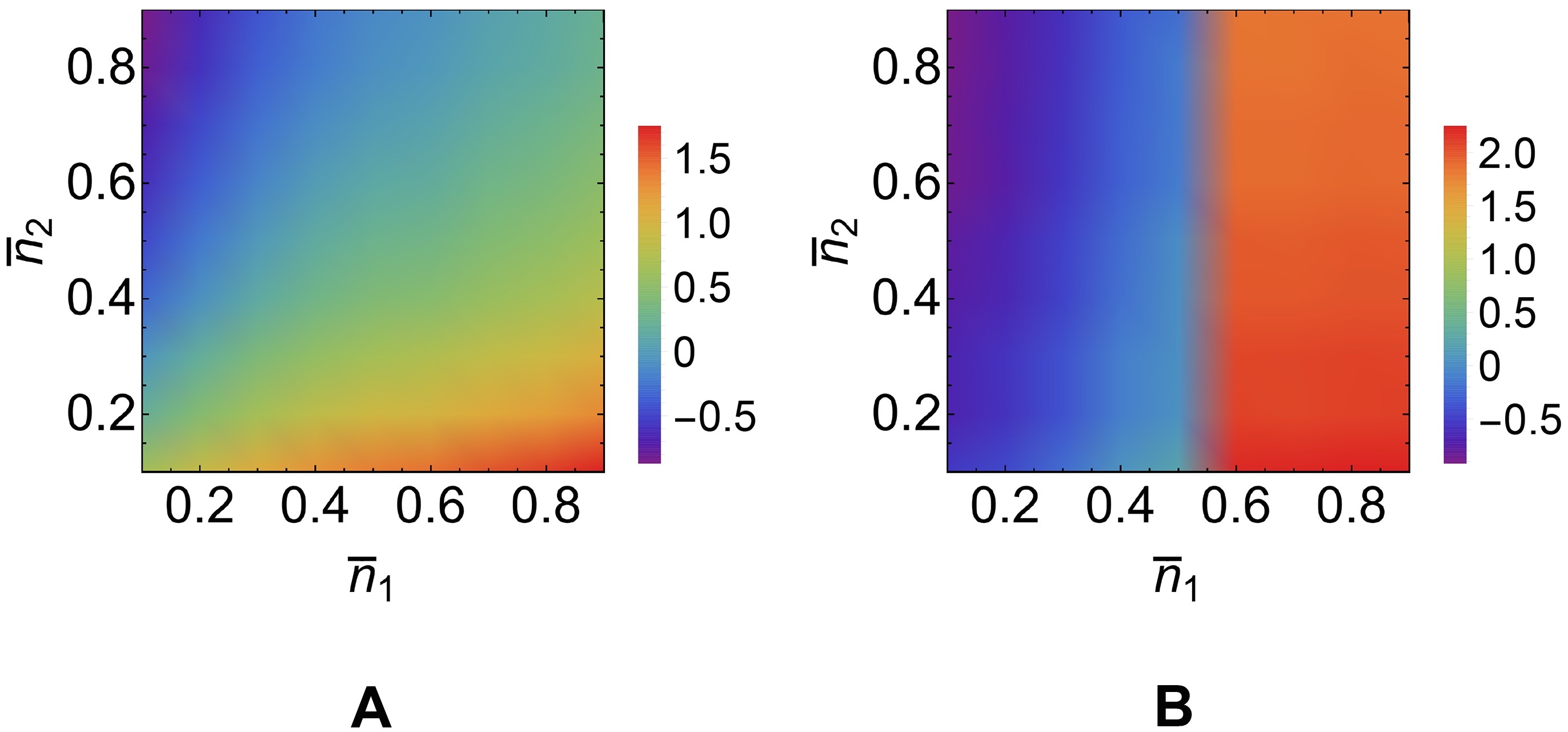}
\end{center}
\caption{
\label{Fig7}
{\bf Dependence of $\Delta I$ on the fractional occupancy of the TF binding site in the optimality range for the post-transcriptional miRNA-mediated channel.} {\bf (A)} Case of targets with weak catalytic degradation (small $\kappa_1$ and $\kappa_2$).
{\bf (B)} Case of a strongly catalytically degraded target (large $\kappa_2$) with a weakly catalytically degraded competitor (small $\kappa_1$). Note that $\bar{n}_\mu\simeq 1$ for (A) while $\bar{n}_\mu\simeq 0.5$ for (B). Values of the kinetic parameters are reported in Table \ref{table1}. 
}
\end{figure}

Transcriptional and post-transcriptional channel capacities are expected to depend on the effective transcription rates of the ceRNAs and the miRNA, described respectively by the quantities $b_i \bar{n}_{i}$ and $\beta \bar{n}_{\mu}$, since these bear a direct impact on the regime (repressed, susceptible or unrepressed) to which each ceRNA belongs. In particular, upon changing the fractional occupancies, situations where miRNA-mediated regulation outperforms TF-based control can occur. Such a possibility might indeed be realized for relatively large values of $k_1^+$, where the miRNA-mediated channel works optimally. This regime is explored in Fig. \ref{Fig7}. 

Remarkably, one finds that post-transcriptional control outperforms transcriptional one when
\begin{enumerate}
\item[(i)] the effective trascription rate of ceRNA$_2$  exceeds that of ceRNA$_1$
 ($ b_2 \bar{n}_2 > b_1 \bar{n}_1$, see in Table \ref{table1}) 
  in presence of weak catalytic degradation (Fig. \ref{Fig7}A);
\item[(ii)] the effective transcription rate  of ceRNA$_1$ is smaller than that of the miRNA ($ b_1 \bar{n}_1 <\beta \bar{n}_\mu$, see in Table \ref{table1}) in presence of strongly catalytically degraded target. 
(see Fig. \ref{Fig7}B).
\end{enumerate}

A more detailed analysis shows that, in these scenarios, whenever $I_\mirna>I_\TF$ the target's derepression size is larger when the signal is processed post-transcriptionally rather than through the direct transcriptional channel, as shown in S3 Figure.
Again, the AOV appears therefore to be a good proxy for the channel capacity.

Previous work had pointed to near-equimolarity of ceRNAs
and miRNAs within a network as one of the central preconditions for significant ceRNA regulation \cite{Figliuzzi,Bosia,Ala}. On the other hand, experiments on liver cells have shown that such conditions may not be met in physiological conditions, where the number of miRNA target sites can vastly exceed that of miRNA molecules \cite{Jens,Denzler}.  Our results suggest that target derepression may be significant even if the competitor is in low copy numbers, provided a certain heterogeneity in kinetic parameters (e.g. for a catalytically degraded target and a stoichiometrically degraded competitor) is present, in line with recent experiments pointing to a key role of parameter diverseness \cite{Bosson}.

\subsubsection* {A sufficiently high degree of target derepression is required for information transmission}

\begin{figure}
\begin{center}
\includegraphics[width=8.5cm]{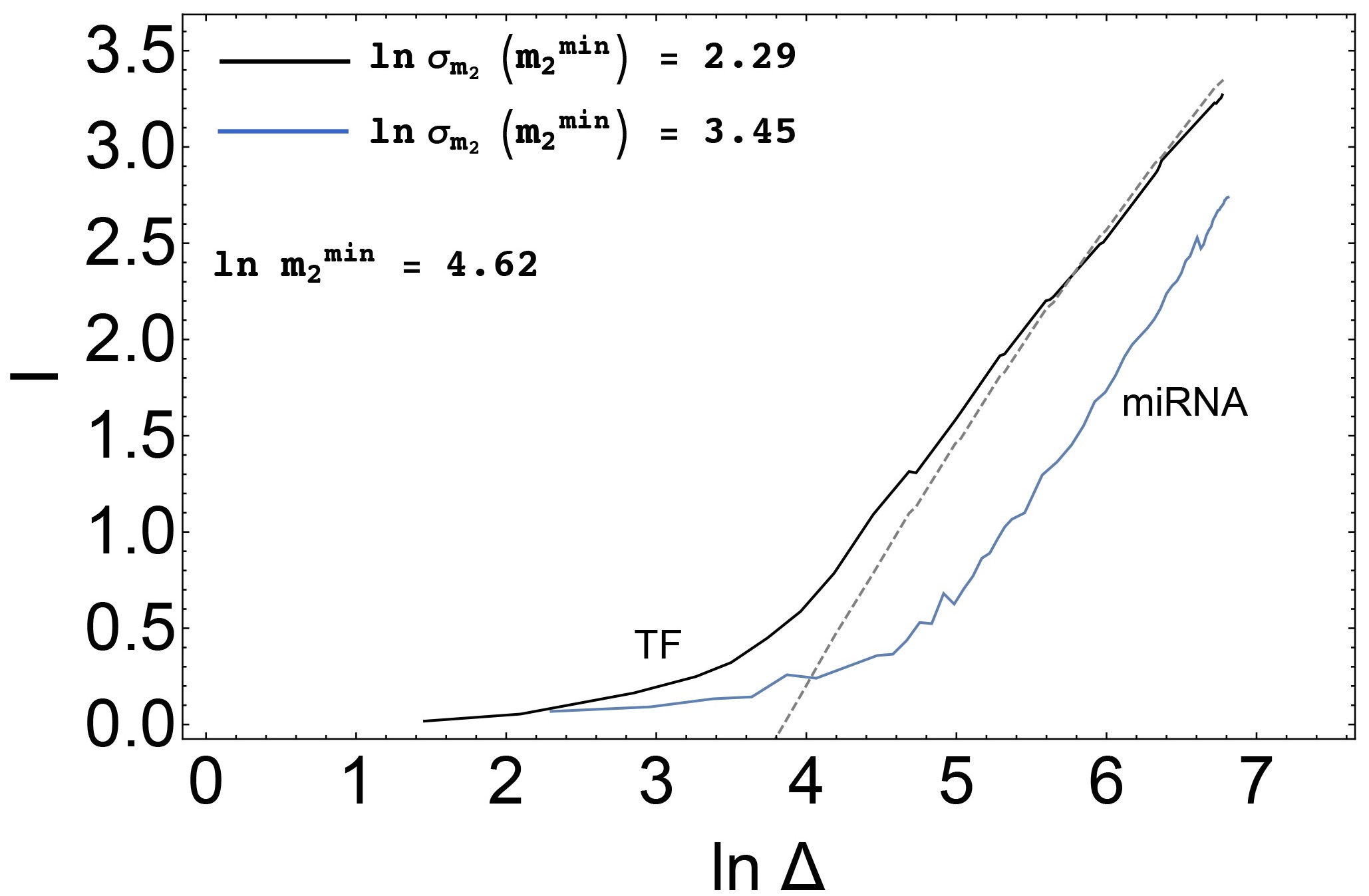}
\end{center}
\caption{
\label{Fig8}
{\bf  Channel capacities as a function of the target's degree of derepression (AOV).} 
The black curve corresponds to direct transcriptional control in absence of miRNAs, while the blue curve describes the behavior of the miRNA-mediated  post-transcriptional channel.
For both channels $\ln m_2^{\min} = 4.62$.
The predicted maximal MI in the Poissonian limit given in S1 Text is shown as a dashed line.
Values of the kinetic parameters are reported in Table \ref{table1}. }
\end{figure}

We have seen that  AOV displays a dependence on kinetic parameters that very closely matches the one found for the channel capacities. In particular, whenever $I_{\mirna}>I_{\TF}$, we found $\Delta_{\mirna}>\Delta_{\TF}$. One may be lead to think that the AOV by itself may fully describe the capacity of miRNA-mediated post-transcriptional information processing and that stochasticity could, to a large degree, be neglected. In some situations, however, the size of the output range is poorly informative about the channel performance. This becomes clear by exploring the relationship between the AOV and the channel capacity at a quantitative level. 
In Fig. \ref{Fig8} the dependence of the optimal MI on the AOV is discussed for two cases:
(i) direct transcriptional regulation by TF in absence of miRNA molecules (black curve), (ii) miRNA-mediated post-transcriptional regulation (blue curve).

In Fig. \ref{Fig8} we indeed see that for large enough AOV, where fluctuations on each level are small compared to $\Delta_\TF$, the TF-channel capacity approaches the theoretical limit of a Poissonian channel worked out in S1 Text.
On the other hand, for small values of $\Delta_\TF$, the capacity of TF-channel becomes vanishingly small.

Similarly in Fig. \ref{Fig8} one can see that $I_\mirna$  increases (roughly) logarithmically with $\Delta_\mirna$ if the latter exceeds a certain threshold, whereas it is essentially zero below this value. In other words, the miRNA-mediated channel is unable to convey any information unless the degree of target derepression is sufficiently high. Hence $I_{\mirna}$ can be close to zero even though $\Delta_{\mirna}$ is finite. Notice that  the derepression threshold for miRNA-mediated information transmission is larger than the corresponding threshold for direct transcriptional information flow.

A close analysis reveals that such a feature is due to the random fluctuations on the lowest expression level of the target, $m_2^{\min}$. Below threshold, in fact, the degree of derepression of the target is smaller than the fluctuations on $m_2^{\min}$ caused by intrinsic noise (reported in Fig. \ref{Fig8}). Therefore the response observed at the level of the output is indistinguishable from the noise. 
In the miRNA-mediated post-transcriptional channel the threshold AOV is larger since noise on $m_2^{\min}$ is higher.
As soon as the AOV overcomes random fluctuations on $m_2^{\min}$, the channel capacities  starts to increase with $\Delta_\mirna$. 
 
\subsubsection*{Noise reduction makes the capacity of the post-transcriptional channel comparable to that of the transcriptional channel in the limit of large populations of weakly interacting miRNAs}

We will now focus specifically on the influence of the miRNA-ceRNA binding component of the intrinsic noise, which will be studied upon varying miRNA-ceRNA binding rates and miRNA population size.  We shall consider two different scenarios, namely (a) direct, TF-based transcriptional regulation in absence of miRNAs (corresponding to $k_i^+ = 0$ for $i = 1, 2$), and (b) indirect post-transcriptional miRNA-mediated regulation (corresponding to non-zero complex binding rates). Kinetic parameters will however be re-scaled as 

\begin{gather}
\delta ~\to~\delta^* = \omega \delta~~,\nonumber \\ 
k_i^+~\to~ (k_i^+)^* = \omega k_i^+~~, \label{eq:rescaling}
\end{gather}
where $\omega>0$ is the re-scaling factor. Note that the above choice does not change $\bar{m}_2$ as long as $k_i^- \bar{c}_i \ll b_i \bar{n}_{i} $ and $(k_i^- + \kappa_i)  \bar{c}_i  \ll \beta \bar{n}_\mu$ (both of which are true in the case we consider). By changing $\omega$ one may therefore characterize the dependence of the channel capacities on the kinetic parameters $\delta$ and $k_i^+$ while keeping the output range approximately fixed, such that $\Delta_\mirna\simeq\Delta_\TF$ for every $\omega$.

\begin{figure}
\begin{center}
\includegraphics[width=8.5cm]{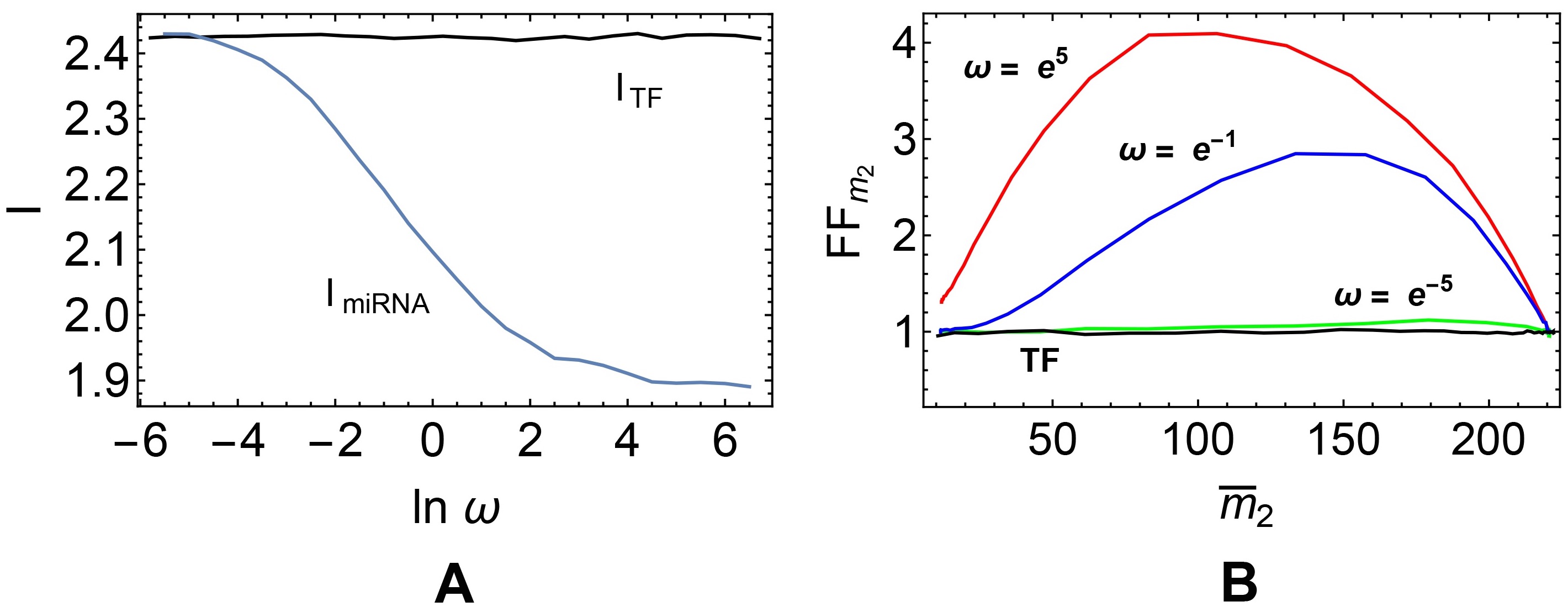}
\end{center}
\caption{
\label{Fig9}
{\bf Comparison of transcriptional and post-transcriptional regulation for a fixed output variation range.}
{\bf (A)} Dependence of $I_\TF$ and $I_{\mirna}$ on $\omega$.
{\bf (B)} FF of $m_2$ at the corresponding steady state. AOV and $m_2^{\min} $ are the same for both channels, namely $\Delta_\mirna = \Delta_\TF = 210 \pm 1$ and $m_2^{\min} = 11 \pm 1$.
Values of the kinetic parameters are reported in Table \ref{table1}.
}
\end{figure}

Fig. \ref{Fig9}A shows how the the channel capacities change with $\omega$. One sees that in general $I_\mirna<I_\TF$, so that at fixed AOV the miRNA-mediated channel typically has smaller capacity than the direct one. However, in the limit of small $\omega$, i.e.  when miRNA populations are sufficiently large and miRNA-ceRNA couplings are weak, the miRNA channel can perform as effectively as a pure transcriptional channel. 
 This result can be understood by examining the noise levels (see Fig. \ref{Fig9}B). 
 For small $\omega$ (or high  $\mu_{0,2}$), in particular, the FF approaches the Poissonian limit since 
 (i) for $\mu \gtrsim \mu_{02}$ relative fluctuations in miRNA levels are small, extra noise coming from molecular titration disappears and the target degradation constant is rescaled as  $d_2\to d_2+k_2^+\bar{\mu}(f_1)$;  (ii) $\mu \ll \mu_{0,2}$ target ceRNA is seen in free regime, where it is not sensitive to miRNA molecules.  In such conditions,  miRNA-mediated noise buffering  \cite{Siciliano} cannot be observed, nevertheless transcriptional and miRNA-mediated regulation are effectively equivalent.

\subsubsection*{Observed heterogeneities in miRNA-target interaction strength may enable regulation by miRNA-mediated cross-talk
} 

The quality of miRNA-target interaction influences the binding kinetics and may be decisive for the activation of the target decay channel \cite{Morozova,Enright,Valencia,BartelDP}.
Estimations of the miRNA-ceRNA binding affinities are experimentally challenging. However,  computational methods predict a considerable degree of heterogeneity across different miRNA-ceRNA pairs \cite{Breda}. Remarkably, for the majority of cases reported in the literature, the predicted miRNA-binding energies of the RNAs on the `input side' of the channel are lower than those of the RNAs on the `output side', in line with the optimal conditions highlighted by our model. For example, binding affinities between the long non-coding RNA linc-MD1 and its regulatory miRNAs (playing a central role in skeletal muscle cell differentiation) are significantly lower than those characterizing miRNA interactions with linc-MD1's competitors, namely the MAML1 and MEF2C mRNAs, as predicted by miRanda algorithm \cite{Cesana,Enright}. Likewise, the circular RNA CDR1 has been found to contain around 70 binding sites with high complementarity for miR-7, corresponding to a strong effective coupling through which it can regulate the expression of miR-7's target genes \cite{Taulli,Memczak,Hansen}. Finally, the high sequence homology of pseudogenes (long non-coding RNA genes developed from protein-coding genes but unable to produce proteins) with their parental gene allows them to compete for a large number of shared miRNAs \cite{Poliseno,Salmena,Karreth}.

Our study also points to the potential relevance of miRNA-ceRNA complex decay channels. It is known that, in case of sufficient complementarity, miRNA molecules can function as siRNAs and cleave their targets after binding them \cite{Valencia,BartelDP}. Such targets decay catalytically and are therefore effectively degraded by the miRNAs. Our model predicts that miRNA-mediated control may be the preferred regulatory mechanism in presence of kinetic heterogeneities at the level of miRNA recycling rates. In particular,  targets that undergo catalytic degradation may be efficiently derepressed by their competitors. This type of scenario has been observed in experiments concerning the bacterial small RNA Qrr \cite{Feng}.  Qrr represses its targets by distinct mechanisms. For instance, luxR is repressed catalytically, luxM stoichiometrically, while luxO is silenced through translational repression. luxM and luxO are however able to derepress LuxR in the presence of Qrr. In the light of our results, these observations may therefore point to a higher than expected role for the ceRNA effect in vivo, especially in cases in which heterogeneities in kinetic parameters are thought to be strong \cite{Bosia,Ala,Wee,Baccarini,Haley}.

\section*{Discussion}

Non-coding RNA molecules, and miRNAs specifically, are increasingly associated to regulatory functions. Besides making it mandatory to characterize the specific role of ncRNAs on a case-by-case basis, especially for situations like disease or differentiation, this fact also raises the question of what ingredients can make miRNAs a preferred tool to regulate the level of a target RNA over, for instance, the target's TF. A possible answer lies in the noise-buffering role that miRNAs can play, which is especially evident in genetic circuitries like incoherent feed-forward loops \cite{Osella,Siciliano}. By reducing relative fluctuations in the output level, miRNAs can confer robustness to gene expression profiles. However the so-called `ceRNA hypothesis' opens the way to the possibility that their regulatory functions are carried out at a broader, though more subtle, level. In short, according to the ceRNA scenario miRNAs can mediate an effective positive interaction between their target RNAs driven by the targets' competition to bind them. In this sense, miRNAs can be seen as a sort of `channel of communication' between RNAs through which RNA levels can be altered and noise can be processed (both buffered and amplified). Previous work \cite{Figliuzzi} has shown that the ceRNA effect may  generate both highly plastic and highly selective ceRNA-ceRNA couplings, thereby representing a potentially powerful mechanism to implement gene regulation at the post-transcriptional level.

Although predicted theoretically, the extent and relevance of ceRNA effect {\it in vivo} is poorly understood. On one hand, considerable evidence points to the ceRNA effect playing a major role in certain dis-regulated or transient cellular states. For instance, it has been shown that the expression of the tumor-suppressor gene PTEN can be regulated by its miRNA-mediated competitors VAPA, CNOT6L, SERINC1 or ZNF460  \cite{TayY}. 
Furthermore, many pseudogenes have been found to compete with their parental genes for a shared pool of common microRNAs, thus regulating their expression as competitive endogenous RNA \cite{Poliseno,Salmena,Karreth,Kalyana-Sundaram} .
Such mechanisms seem to be of particular relevance in cancer. For instance, murine models engineered to overexpress the pseudogenes of the proto-oncogene BRAF develop an aggressive malignancy resembling human B cell lymphoma since, by functioning as ceRNAs, they elevate BRAF expression both {\it in vitro} and {\it in vivo} \cite{Karreth}.
Likewise, the long noncoding RNA linc-MD1 has been shown to regulate the skeletal muscle cell differentiation clock by sponging miRNAs from its competitors,  thereby enacting a ceRNA mechanism. In particular, MAML1 and MEF2C (coding for transcription factors that activate muscle-specific gene expression) compete with linc-MD1 for miR-133 and miR-135 respectively \cite{Cesana}. Taken together, the available evidence indicates that miRNA activity depends on the miRNA:target ratio, on miRNA target site abundance and on miRNA binding affinities. Further analyses of high throughput datasets confirm this observation \cite{Bosson,Denzler}. One may therefore question how said factors may influence miRNA-mediated post-transcriptional control.

The problem however arises of quantifying the degree of control that can be exerted through miRNAs. Taking the `channel' analogy more strictly (as done before for simpler regulatory elements \cite{Tkacik,review,Bialek}), one may resort to information theoretic concepts and tools to characterize precisely how well miRNAs can process fluctuations coming from the modulator nodes and transfer them to the target nodes. This issue goes beyond noise buffering, specifically including the ability to respond to large changes in the mean levels as well as to changes in the structure of fluctuations. As the properties of a channel are conveniently encoded in the mutual information between the input and output nodes, asking how well a channel can function amounts to asking what is the channel's capacity, i.e. the maximum value of the input-output (or modulator-target) mutual information achievable through that channel. This work aimed precisely at quantifying the effectiveness of microRNA-mediated post-transcriptional control of gene expression by computing the capacity of the corresponding regulatory channel and comparing it to that of direct, TF-driven transcriptional regulation. 
Evidently, multiple factors can influence the flow of information across nodes in a biochemical network, starting from the intrinsic noisiness of each reaction step. 

Our basic challenge was therefore understanding in which circumstances miRNA-mediated control can outperform the TF-based one, thereby obtaining insight on why the ceRNA effect appears to be so often employed by cells in situations where accurate tuning and/or shifts of expression levels are required. 
We have therefore considered, along the lines of \cite{Figliuzzi,MFigliuzzi,Bosia,CarlaB,Mehta,Yuan}, a mathematical model of the ceRNA effect and characterized its steady state in terms of both mean molecular levels and regulatory capacities of the miRNA-mediated and TF-based channels via stochastic simulations.

We have first considered how the two channels process inputs (the TF levels $f_1$ and $f_2$) that vary in the same range. We have shown that, while the capacity of the TF-channel depends monotonously on each miRNA-ceRNA binding rate and is largest when the target is unrepressed by miRNAs (as might have been expected), the capacity of the post-transcriptional channel is maximal in a specific range of values of the miRNA-ceRNA binding rates. In agreement with \cite{Yuan}, we found that miRNA-channel's efficiency is tunable to optimality by the binding kinetics. 
Furthermore, our model suggests that both capacities decrease as the miRNA recycling rates increase, confirming previous indications obtained by different analytical techniques \cite{Figliuzzi}. Consistently with the scenario observed experimentally for the bacterial small RNA Qrr \cite{Feng}, our model finally suggests that catalytically regulated targets are weakly capable of competing for miRNAs but might be significantly derepressed by their competitors.

In addition,  post-transcriptional miRNA-mediated information processing was shown to be characterized by a threshold behaviour as a function of the AOV. In other terms, no information can be transmitted across the channel unless the target's degree of derepression is sufficiently large. This implies that the regulatory effectiveness of the channel is well encoded by the degree of target derepression when the latter is sufficiently high, in which case it is possible to identify regimes in which post-transcriptional regulation is more accurate than transcriptional control.

To get a deeper insight on the origin of the observed threshold behaviour one must however go beyond the AOV and consider more carefully how the miRNA-ceRNA binding noise affects the overall picture. After showing that miRNA-ceRNA binding noise is indeed at the origin of the threshold behaviour that limits the miRNA-channel capacity, we have uncovered the rather remarkable property that in presence of large but weakly interacting miRNA populations the ceRNA effect can regulate gene expression as effectively as the target's modulator node itself. 

The present work has focused on a small genetic circuit made up of a single miRNA species and two target RNA species at steady state. Previous work has however shown that cross-talk is possible even during transients \cite{MFigliuzzi}. Going beyond stationarity is therefore likely to bring to light new scenarios where miRNA-mediated regulation plays possibly a yet more prominent role.

On the other hand in a typical eukaryotic cell there are thousands ceRNAs, hundreds miRNAs and a rich structure of conserved targeting patterns \cite{Obermayer}. Moreover, cells might be interested in tightly controlling not only each output individually but also particular combinations of output levels (which might be required e.g. for the efficient operation of metabolic pathways). In such a scenario, miRNA-mediated control could represent a powerful mechanism to increase robustness and flexibility in specific directions of the output space. In view of this, it would be important to consider a more general multi-source network coding problem in which a large number of transcription processes are seen as mutually independent information sources, and each of the information sources is multicast to sets of output nodes through the effective network of miRNA-mediated cross-talk interactions. The information-theoretic scheme employed in this work is easily generalized to deal with more complex networked situations. Novel insight might finally shed light on the partly controversial picture unveiled by recent experiments addressing the relevance of the ceRNA effect {\it in vivo} \cite{Denzler,Cesana,Karreth,Ebert,Bosson,Memczak}.

\section*{Materials and methods}
\label{MM}
% Include only the SI item label in the subsection heading. Use the \nameref{label} command to cite SI items in the text.

\subsection*{Gillespie algorithm}
\label{GA}

Numerical simulations have been carried out using the Gillespie algorithm (GA), a standard stochastic method to analyze the time evolution of a system of chemical reactions which is exact for  spatially homogeneous systems \cite{Gillespie}. In short, based on the reaction rates, GA calculates when the next reaction will occur and what reaction it will be, and then modifies the amount of each molecular species in the system according to the process that took place. If we denote the probability of reaction $r$ to occur in the time interval $(\tau, \tau + d \tau)$ by $P(r,\tau) d\tau$, the algorithm proceeds through the following steps: 
\begin{enumerate}
\item[(1)] Initiate the number of reactants in the system and the termination time;
\item[(2)] Generate a random pair $(r,\tau)$ according to $P(r, \tau)$;
\item[(3)] Using the pair $(r,\tau)$ just generated, advance time by $\tau$ and change number of species involved in reaction $r$ accordingly;
\item[(4)] Read out the molecular population values. If the termination time is reached, stop the simulations, otherwise return to Step 2.
\end{enumerate}
After a long run, independently on the initial setup, the system of chemical reactants will come to the equilibrium state. 

~

\subsection*{Linear noise approximation}
\label{LNA}

The mathematical model of ceRNA competition can be solved numerically in the so-called linear noise approximation. Letting $\mathbf{x} = (m_1, m_2, \mu, c_1, c_2) $ stand for the vector of molecular levels, the kinetic mass action equations (\ref{eq:SDEeqs}) can be re-cast in compact form as
\begin{equation}
\frac{d\mathbf{x}}{dt}  = \mathbf{g}(\mathbf{x}) + \boldsymbol{\eta}~~,
\end{equation}
where the vector $\mathbf{g}$ encodes for the deterministic part of the dynamics, while the vector $\boldsymbol{\eta}$ represents the aggregate noise terms. Each element of $\boldsymbol{\eta}$ has zero mean, and we shall denote its correlations by $\avg{\eta_a(t) \eta_b (t')} = \Gamma_{ab} \delta (t-t')$.  Denoting by $\bar{\mathbf{x}}$ the steady state, small deviations from it (i.e. $\delta\mathbf{x}=\mathbf{x}(t)-\bar{\mathbf{x}}$) relax, in the linear regime, according to
\begin{equation}
\frac{d }{d t} \delta \mathbf{x} = \mathbf{A} \delta \mathbf{x} + \boldsymbol{\eta}~~,
\end{equation}
where $\mathbf{A}= \frac{d\mathbf{g}}{d\mathbf{x}}\big|_{\mathbf{x}=\bar{\mathbf{x}}}$. In this approximation, the correlation matrix $C_{ab} = \avg{\delta x_a \delta x_b}$ is given by \cite{Swain}
\begin{equation}
C_{ab} =- \sum_{p,q,r,s} B_{ap}B_{br}\frac{\Gamma_{qs}}{\lambda_p+\lambda_r}(B^{-1})_{pq}(B^{-1})_{rs}~~,
\label{corr}
\end{equation}
where $\lambda$'s and $B$'s are, respectively, eigenvalues and eigenvectors of the matrix $\mathbf{A}$.

\subsection*{Model parameters}
\label{Parameters}
\addcontentsline{toc}{subsection}{Model parameters}

Table \ref{table1} reports the values of the kinetic parameters (or of their range of variability) employed in the different figures. As $k_1^+, k_2^+, \kappa_1, \kappa_2, b_1, b_2, b_\mu$ and $d_\mu$ are varied in wide ranges in order to test how channel capacities depend on them, our choice was mainly guided by the need to focus the analysis on regimes where ceRNA cross-talk is established (so that the miRNA-mediated channel can actually convey information).

\section*{Author Contributions}

Conceived and designed the experiments: MF EM ADM. Performed the experiments: AM. Analyzed the data: AM MF EM ADM. Contributed reagents/materials/analysis tools: AM MF EM ADM. Wrote the paper: AM MF EM ADM.

%\begin{comment}
\begin{widetext}
%   \clearpage% Flush earlier floats (otherwise order might not be correct)
%    \thispagestyle{empty}% empty page style (?)
%    \begin{landscape}% Landscape page
        \centering % Center table
\begin{tabular}[t]{c c  c c c c c c c c}
\rowcolor{gray!20}
Parameter &  Fig. \ref{Fig2}A, B   & Fig. \ref{Fig5}A, B, C  & Fig. \ref{Fig6}A, B, C  & Fig. \ref{Fig7}A  & Fig. \ref{Fig7}B  & Fig. \ref{Fig8} & Fig. \ref{Fig9}\\
\hline
\hline
$b_1$ [molecule min$^{-1}$]		&	120	        & 100	&		100	&	100	&	100 & 	  \pbox{20cm}{  [1, 147] (miRNA); \\ 0 (TF)  }    &   \pbox{20cm}{ 90 (miRNA);  \\ 0 (TF)}	 \\
\rowcolor{gray!15}
$b_2$ [molecule min$^{-1}$]		&	100	        & 100	&	100	&	100	&	100 &    \pbox{20cm}{  $110.2$ (miRNA); \\ $[10, 98]$ (TF) }   & 	  \pbox{20cm}{22.3 (miRNA);\\ 22 (TF)}  \\ 
$\beta$	[molecule min$^{-1}$]	&	100	        & 100	&	100	&	100	& 	100 &	 \pbox{20cm}{  100 (miRNA); \\ 0 (TF)} &	
\pbox{20cm}{  80 (miRNA);\\ 0 (TF)} \\
\rowcolor{gray!15}
$d_1$[min$^{-1}$]		&	0.1		& 0.1	&	0.1	&	0.1		&	 0.1 &	 0.1 &	 0.1	\\
$d_2$ [min$^{-1}$]		&	0.1		& 0.1	&	0.1	&	0.1		&	 0.1 &	 0.1 &	 0.1	\\
\rowcolor{gray!15}
$\delta$ [min$^{-1}$]		&	0.1		& 0.1	&	0.1	&	0.1		&	 0.1 &	 0.1  & 0.1 $\omega$	\\
$\sigma_1$ [min$^{-1}$]	&	0.5		&	1	&	1	&	1		&	1	&	1	& 1\\
\rowcolor{gray!15}
$\sigma_2$ [min$^{-1}$]	&	1		&	1	&	1	&	1		&	1	&	1	&	1\\
$k_1^+$	[molecule$^{-1}$ min$^{-1}$]	&	0.002 	&	  $[e^{-10}, e^2]$  		&	$e^{2}$	& $e^{2}$  &	$e^{2}$   &   \pbox{20cm}{$e^3 $ (miRNA); \\ 0 (TF)  } &	 \pbox{20cm}{ $e^{3.39}\omega$ (miRNA);\\ 0 (TF)}  	 \\
\rowcolor{gray!15}
$k_2^+$	[molecule$^{-1}$ min$^{-1}$]	&	0.001	 &	 $[e^{-10}, e^2]$ 	&	$e^{-3}$		&	$e^{-3}$ &	$e^{-3}$  &	  \pbox{20cm}{$e^{0} $(miRNA); \\ 0 (TF)  } & \pbox{20cm}{ $e^{-5.77}\omega$ (miRNA); \\0 (TF) }\\  
$k_1^-$	[min$^{-1}$]	&	0.0005	&	0.001	&	0.001	&	0.001	&	0.001	& 0.001	 	&	0.001		\\
\rowcolor{gray!15}
$k_2^-$	[min$^{-1}$]	&	0.001	&	0.001	&	0.001	&	0.001	&	0.001	&	0.001		&	0.001		\\
$\kappa_1$ [min$^{-1}$]	&	0.0003	&0.002		&	 $[e^{-5}, e^5]$	&		0.002 &	$e^{-8}$	 & 0.002  &	0.002\\
\rowcolor{gray!15}
$\kappa_2$ [min$^{-1}$]	&	0.0004	&0.002	&	 $[e^{-5}, e^5]$		&	0.002 &	$e^{5}$		 &   0.002  &	0.002\\
$\bar{n}_{1}$ [adim.]		&	 (0,0.99)   &   \pbox{20cm}{(0,0.99) (miRNA);\\ 0.99(TF) } &	 \pbox{20cm}{(0, 0.99) (miRNA);\\ 0.21(TF) } 	&	(0.1,0.9)	& (0.1,0.9) &  (0, 0.99) (miRNA) & (0, 0.99) (miRNA)	\\
\rowcolor{gray!15}
$\bar{n}_{2}$ [adim.]	&	0.71		& 
\pbox{20cm}{0.99 (miRNA);\\  (0,0.99) (TF)}	&\pbox{20cm}{0.21 (miRNA);\\ (0, 0.99) (TF)} 	&	(0.1,0.9)	& (0.1,0.9) &  \pbox{20cm}{0.99 (miRNA); \\ (0, 0.99) (TF)}  &  \pbox{20cm}{0.99 (miRNA); \\ (0, 0.99) (TF)}	\\
$\bar{n}_\mu$ [adim.] &	0.9		&	0.99 & 0.99	&	0.99	&			0.5  & 0.99 & 0.99		\\
\rowcolor{gray!15}
$h$ [adim.]		&	5	&	5	& 5 &	5		&	5	&  5	& 5	\\
$k_{\out}/k_{\inn}$ [molecules$^h$]	&	63300	&	63300	&	63300	&	63300&	63300	&	63300	&	63300	\\
\rowcolor{gray!15}
$f^{\max}$ [molecules]		&	30	&	30	&	30	&	30	&	30	&	30	&	30 		\\
\hline
\end{tabular}
\bigskip\
        \captionof{table}{
\label{table1}
{\bf Parameters used to obtain each figure.} 
When the parameter was varied over a range of values, the range is shown in brackets. The parameter $\omega$ appearing in the last column denotes the re-scaling factor defined in Eq. \eqref{eq:rescaling}, while the channel to which parameters refer is specified in brackets. Values of $f_1, f_2 $ and $ f_\mu$ are mapped to those of $\bar{n}_1, \bar{n}_2, \bar{n}_\mu $ via the steady state conditions, Eq. \eqref{eq:nll}.
}% Add 'table' caption
%    \end{landscape}
\end{widetext}  
%  \clearpage% Flush page

%\end{comment}

\clearpage

\begin{widetext}

\section*{Supporting Information}
%\addcontentsline{toc}{section}{Supporting Information}
%\vspace{5 mm}
%\indent

{\bf S1 Text. Capacity of a Poissonian channel.} In the absence of miRNA molecules, fluctuations in TF-channel are Poissonian (i.e. the corresponding FF is one). In such a case, the channel capacity (see Eq. (\ref{eq:Iopt})) is given by
\[
I_\opt=\log_2\frac{1}{\sqrt{2\pi e}}\int_{f_j^{\min}}^{f_j^{\max}}\frac{1}{\sqrt{\bar{m}_2}}\frac{\partial\bar{m}_2}{\partial f_j} df_j=\log_2\frac{1}{\sqrt{2\pi e}}\int_{m_2^{\min}}^{m_2^{\max}}\frac{d\bar{m}_2}{\sqrt{\bar{m}_2}}=\log_2\left[\sqrt{m_2^{\max}}-\sqrt{m_2^{\min}}\right]-\frac{1}{2}\log_2\frac{\pi e}{2}
\]
where $m_2^{\min}=\bar{m}_2(f_j^{\min})$ and $m_2^{\max}=\bar{m}_2(f_j^{\max})$ are the minimum and maximum expression levels of the target, respectively.

\vspace{8mm}

\begin{figure*}[h!]
\begin{center}
\includegraphics[width=13cm]{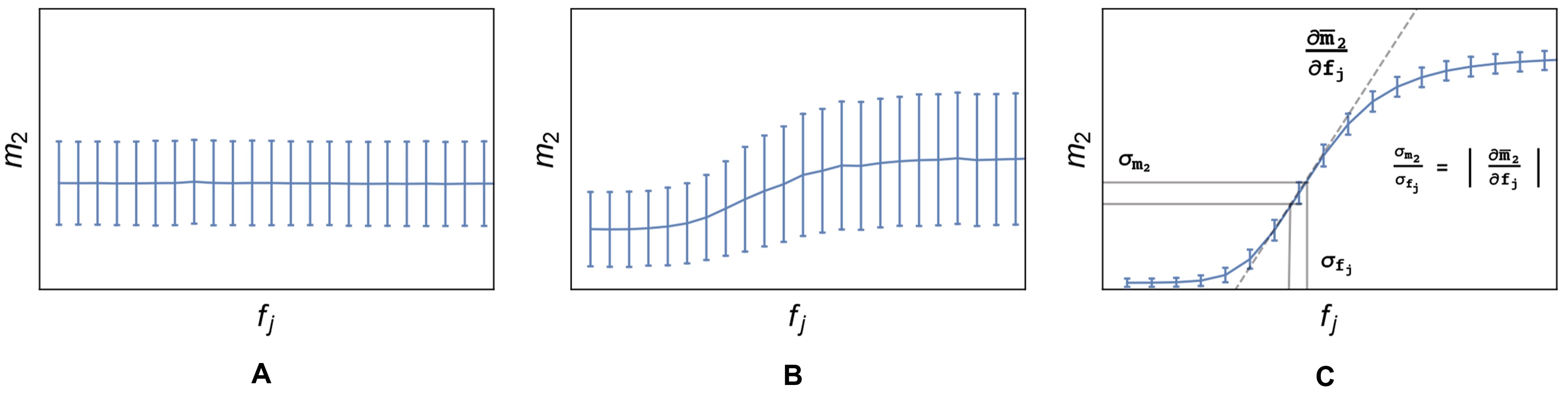}
\end{center}
\caption{
{\bf S1 Figure. Information transmission and noise.} Qualitative depiction of how noise affects information flow. Three different situations for the input-output curve $m_2$ (target) vs $f_j$ (modulator) are shown, namely, long
 {\bf (A)} target expression level independent of the modulator; 
 {\bf (B)} derepressed target, with large fluctuations;
 {\bf (C)} derepressed target, with small fluctuations. In the first two cases, little or no information can be transmitted from modulator to target (hence the regulatory effectiveness is severely limited), as either the target is insensitive to changes in modulator levels, or its response is strongly hindered by noise. On the other hand, in case (C) information will be transmitted, since modulating the input one can clearly distinguish different output levels. The number of distinguishable levels is linked to the local slope of the input/output curve, as shown mathematically in Eqs. (\ref{eq:zzeta}) and (\ref{eq:Iopt}) and is mainly limited by the noise strength. In the limit of vanishing noise, when the input-output relationship becomes deterministic, the mutual information between $m_2$ and $f_j$ diverges.
}
\end{figure*}

\begin{figure*}[h!]
\begin{center}
\includegraphics[width=13cm]{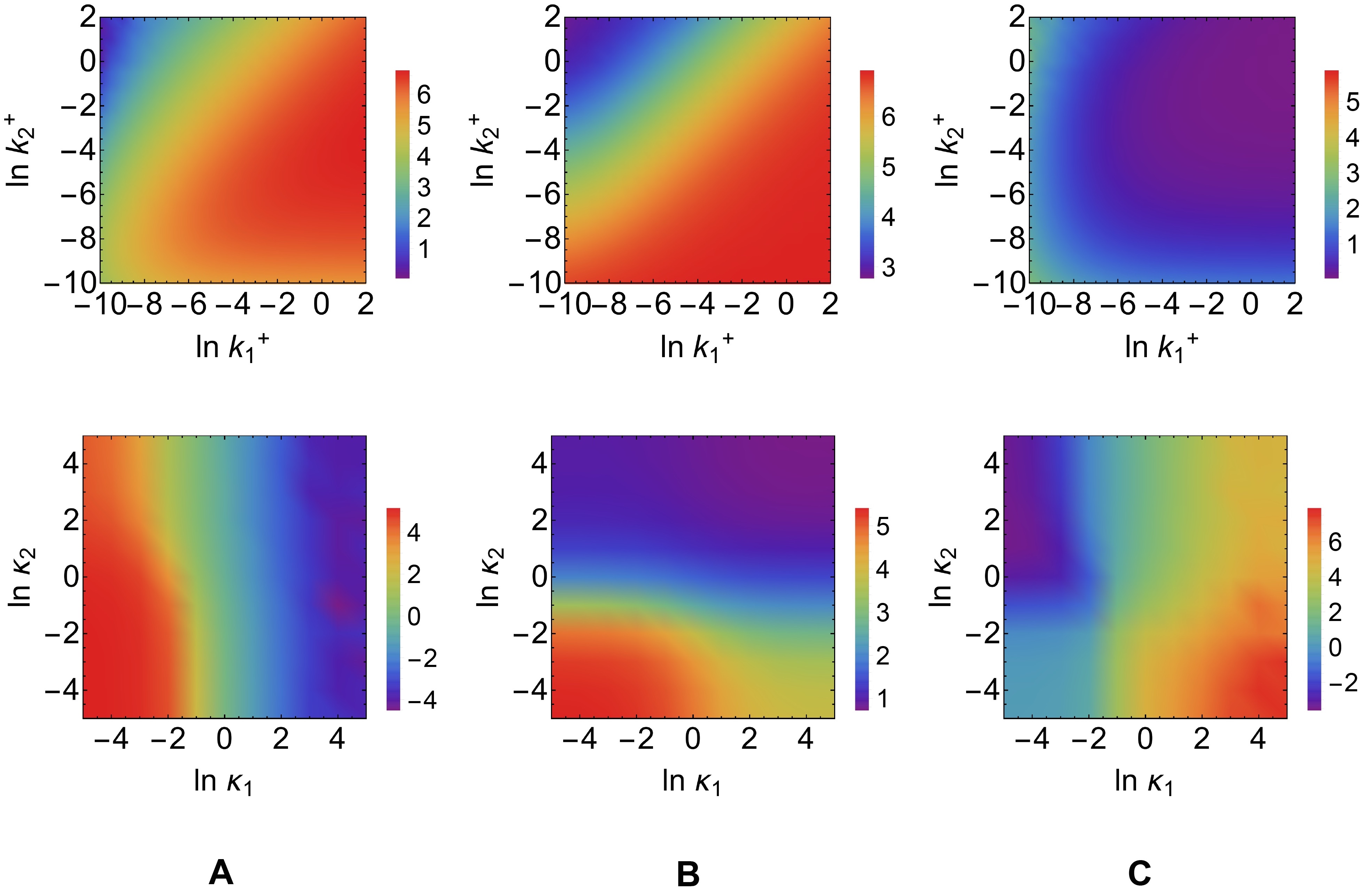}
\end{center}
\caption{
{\bf S2 Figure. Dependence of the AOV on kinetic parameters.}
{\bf (A)} $\ln \Delta_\mirna$;
{\bf (B)} $\ln \Delta_\TF$ ;
{\bf (C)} $\ln \Delta_\TF$ - ln $\Delta_\mirna$ .
Values of the kinetic parameters are as in Fig. \ref{Fig5} for the panels in the top row and as in Fig. \ref{Fig6} for the panels in the bottom row.
}
\end{figure*}

\begin{figure*}[h!]
\begin{center}
\includegraphics[width=9cm]{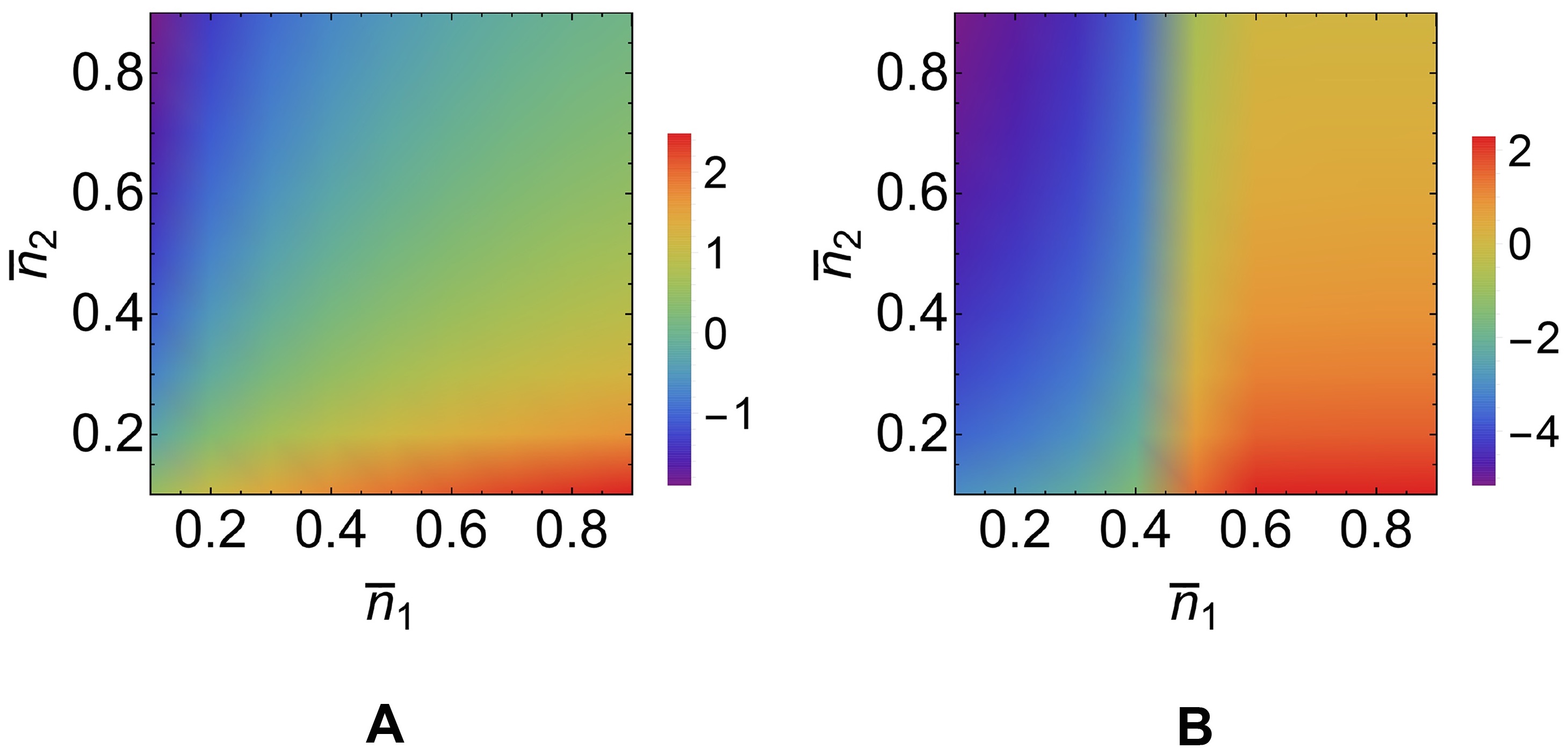}
\end{center}
\caption{
{\bf S3 Figure. Dependence of $\ln \Delta_\TF - \ln \Delta_\mirna$ on the fractional occupancy of the TF binding site.} Values of the kinetic parameters are as in Fig. \ref{Fig7}A for panel {\bf (A)}, and as in Fig. \ref{Fig7}B for panel {\bf (B)}.
}
\end{figure*}

{\bf S1 Dataset. Datasets used in the figures} are available as Supporting Information from \href{http://journals.plos.org/ploscompbiol/article?id=10.1371/journal.pcbi.1004715}{\underline{this link}}.

\end{widetext}

\end{document}